\documentclass[preprint2]{aastex}
\def\Msol{\thinspace\hbox{$\hbox{M}_{\odot}$}}

\def\a4{\hsize 17.0cm \vsize 25.cm}
\newcommand{\der}[2]  { \frac{{\rm d}#1}{{\rm d}#2} }

\newcommand{\dif}     {{\rm d}}

\shorttitle{Winds driven by super-star clusters}
\shortauthors{Silich et al.}

\begin{document}

\title{Winds driven by super-star clusters: The self-consistent 
radiative solution}

\author{Sergey Silich}
\affil{Instituto Nacional de Astrof\'\i sica Optica y
Electr\'onica, AP 51, 72000 Puebla, M\'exico; silich@inaoep.mx}

\author{Guillermo  Tenorio-Tagle}
\affil{Instituto Nacional de Astrof\'\i sica Optica y
Electr\'onica, AP 51, 72000 Puebla, M\'exico; gtt@inaoep.mx}

\and

\author{Ary Rodr\'{\i}guez-Gonz\'alez}
\affil{Instituto Nacional de Astrof\'\i sica Optica y
Electr\'onica, AP 51, 72000 Puebla, M\'exico; ary@inaoep.mx}

\begin{abstract}
Here we present a self-consistent stationary solution for spherically 
symmetric winds driven by massive star clusters under the impact of
radiative cooling. We demonstrate that cooling may modify drastically 
the distribution of temperature if the rate of injected energy 
approaches a critical value. We also prove that the stationary 
wind solution does not exist whenever the energy radiated away at 
the star cluster center exceeds $\sim 30\%$ of the energy deposition 
rate. Finally we thoroughly discuss the expected appearance of 
super-star cluster winds in the X-ray and visible line regimes. 
The three solutions here found: the quasi-adiabatic, the strongly
radiative wind and the inhibited stationary solution, are then
compared to the winds from Arches cluster, NGC 4303 central cluster
and to the supernebula in NGC 5253.
\end{abstract}


\section{Introduction}

In the stationary solution for spherically symmetric winds
(Chevalier and Clegg 1985; hereafter referred to as CC85) as well 
as in the former approach of Holzer and Axford (1970) and in the
more recent numerical calculations of Cant{\'o} et al. (2000) and 
Raga et al. (2001) the flow has been assumed to be adiabatic and thus 
predicts a very extended X-ray envelope around the sources.
The impact of cooling on the stationary wind solution, was 
discussed by Silich et al. (2003, hereafter referred to as Paper I) 
for winds driven by powerful and compact stellar clusters, and by Wang
(1995) for gas outflows from galaxies. Winds driven by compact star 
clusters establish a temperature distribution radically different from 
that predicted by the adiabatic solution, bringing the X-ray emitting
boundary much closer to the star cluster surface. However, in none
of the above studies, the effects of radiative cooling within the 
star forming volume itself were taken into consideration.

Here we present a self-consistent semi-analytical model 
of stationary winds driven by massive stellar clusters taking full
account of radiative cooling (see sections 2 and 3). We first 
discuss how to find proper wind central values and then use them to 
integrate numerically the basic equations. We also indicate the 
threshold value of the energy deposition rate above which a 
stationary solution is inhibited. In sections 4 and 5, the three
regimes found when radiative cooling is considered: the
quasi-adiabatic, the strongly radiative wind and the inhibited
stationary wind, are then compared to well observed examples. Our
conclusions are given in section 6.

\section{The adiabatic solution}

Following CC85, Cant{\'o} et al. (2000) and Raga et al. (2001), we assume 
that within a star cluster,  within the volume of radius $R_{sc}$, the matter 
ejected by stellar winds and supernova explosions is fully thermalized 
via random interactions. This generates the large central overpressure 
that continuously accelerates the ejected gas and eventually blows it 
out of the star cluster volume. There are three star cluster
parameters which together define the hydrodynamical properties of 
the resultant wind outflow (or the run of density, temperature 
and expansion velocity, which asymptotically approach
$\rho_w \sim r^{-2}$, $T_w \sim r^{-4/3}$ $u_w \approx V_{\infty A}$). 
The three parameters are: the total energy (${\dot E}_{sc}$) and mass 
(${\dot M}_{sc}$) deposition rates and the actual size of the volume 
that encloses the star cluster ($R_{sc}$). 
The total mass and energy deposition rates also define the wind
terminal velocity $V_{\infty A} = \sqrt{2 {\dot E}_{sc} / {\dot M}_{sc}}$.

In the adiabatic case there is an analytic solution and thus one can 
derive the wind central density, pressure and temperature (see
Cant{\'o} et al., 2000) if the above parameters are known:
%---------------------------------------------------------------
\begin{eqnarray}
      \label{eq.0a}
      & & \hspace{-1.0cm}
\rho_c = \frac{{\dot M}_{sc}}{4 \pi B R_{sc}^2 V_{\infty A}} ,
      \\[0.2cm] \label{eq.0b}
      & & \hspace{-1.0cm}
P_{c} = \frac{\gamma - 1}{2 \gamma} \frac{{\dot M}_{sc} V_{\infty A}}
        {4 \pi B R_{sc}^2} , 
      \\[0.2cm] \label{eq.0c}
      & & \hspace{-1.0cm}
T_c = \frac{\gamma-1}{\gamma} \frac{\mu}{k} \frac{q_e}{q_m} , 
%c_c  = \sqrt{\frac{\gamma - 1}{2}} V_{\infty A} ,
\end{eqnarray}
%-------------------------------------------------------------
where $
B     = \left(\frac{\gamma-1}{\gamma+1}\right)^{1/2}
        \left(\frac{\gamma+1}{6\gamma+2}\right)^
        {(3\gamma+1)/(5\gamma+1)}$ ,
$q_e$ and $q_m$ are the energy and mass deposition rates per unit 
volume ($q_e = 3 {\dot E}_{sc}/4 \pi R_{sc}^3$; $q_m = 3 {\dot M}_{sc}/4 \pi
R_{sc}^3$), $\gamma$ is the ratio of specific heats, $\mu$ is the mean 
mass per particle and $k$ is the Boltzmann constant.
Using these initial values one can solve the stationary 
wind equations numerically and reproduce the analytic solution
throughout the space volume. The relevant equations are:
%---------------------------------------------------------------
\begin{eqnarray}
      \label{eq.1d}
      & & \hspace{-1.0cm}
\frac{1}{r^2} \der{}{r}\left(\rho_w u_w r^2\right) = q_m ,
      \\[0.2cm]
      \label{eq.1e}
      & & \hspace{-1.0cm}
\rho_w u_w \der{u_u}{r} = - \der{P_w}{r} - q_m u_w,
      \\[0.2cm]
     \label{eq.1f}
      & & \hspace{-1.0cm}
\frac{1}{r^2} \der{}{r}{\left[\rho_w u_w r^2 \left(\frac{u_w^2}{2} +
\frac{\gamma}{\gamma - 1} \frac{P_w}{\rho}\right)\right]} = q_e - Q,
\end{eqnarray}
%-------------------------------------------------------------
In equations (\ref{eq.1d}-\ref{eq.1f}) $r$ is the spherical
radius and $Q$ is the cooling rate, assumed equal to zero in CC85, 
Cant\'o et al. (2000) and Raga et al. (2001).

Within the central volume, temperature and density present  almost 
homogeneous values, where\-as the expansion velocity grows almost
linearly from 0 km s$^{-1}$ at the center, to the sound 
speed at the cluster radius, $r = R_{sc}$.
There is then a rapid evolution as matter streams away from 
the star cluster. The flow accelerates rapidly when approaching the  
sonic point  and the wind temperature and density begin to deviate
from their central quasi-homogeneous distributions. At large radius 
the resultant wind parameters rapidly approach their asymptotic values 
%$V_w \to $ $V_{\infty}$, $\rho_w \sim $ $r^{-2}$, $T_w \sim $ $r^{-4/3}$ 
(see CC85 and paper I).

\section{The radiative solution}

Due to the highly nonlinear character of the cooling function, the analytic
approach is not valid in the general case that includes radiative cooling,
and thus one needs to perform a numerical integration. 
However in such a case, the star cluster parameters (${\dot E}_{sc}$, ${\dot
M}_{sc}$ and $R_{sc}$) do not define the wind central temperature 
and density and the problem arises: how to solve equations 
(\ref{eq.1d}-\ref{eq.1f}) if neither the initial nor the boundary 
conditions are known? 

To solve the problem we re-write equations (\ref{eq.1d}-\ref{eq.1f}) and 
obtain within the star cluster radius ($r \le R_{sc}$) 
%---------------------------------------------------------------
\begin{eqnarray}
      \label{eq.2a}
      & & \hspace{-1.1cm} 
\der{u_w}{r}  = \frac{1}{\rho_w} \frac{(\gamma-1)(q_e - Q) + 
              q_m (\frac{\gamma+1}{2}u_w^2 - \frac{2}{3} c_s^2)}
              {c_s^2 - u_w^2} ,
      \\[0.2cm]     \label{eq.2b}
      & & \hspace{-1.1cm}
\der{P_w}{r} = - q_m \left(\frac{r}{3} \der{u_w}{r} + u_w\right),
      \\[0.2cm] \label{eq.2c}
      & & \hspace{-1.1cm}
\rho_w = \frac{q_m r}{3 u_w} ,
\end{eqnarray}
%------------------------------------------------------------- 
and for $r >$ $R_{sc}$ 
%---------------------------------------------------------------
\begin{eqnarray}
      \label{eq.3a}
      & &         \hspace{-3.0cm}
\der{u_w}{r}  = \frac{1}{\rho_w} \frac{(\gamma-1) r Q + 
              2 \gamma u_w P_w}{r (u_w^2 - c_s^2)} ,
      \\[0.2cm]   \label{eq.3b}
      & & \hspace{-3.0cm}
\der{P_w}{r} = - \frac{{\dot M}_{sc}}{4 \pi r^2} \der{u_w}{r} ,
      \\[0.2cm]   \label{eq.3c}
      & & \hspace{-3.0cm}
\rho_w = \frac{{\dot M}_{sc}}{4 \pi u_w r^2} .
\end{eqnarray}
%-------------------------------------------------------------
In equations (\ref{eq.2a}-\ref{eq.3c}) $P_w$ is the wind thermal 
pressure, $c_s = (\gamma P/\rho)^{1/2}$ is the sound speed,  
$Q = n_w^2 \Lambda$, $n_w$ is the wind atomic number density and 
$\Lambda(Z,T)$ is the cooling function (a function of metallicity
and temperature, see Raymond et al. 1976).

It is easy to prove that the derivative of the expansion velocity 
is positive throughout the space volume, only if the sonic point 
is at the star cluster surface. Indeed, if the expansion velocity becomes 
supersonic at $r < R_{sc}$, the right-hand side of equation
(\ref{eq.2a}) and $\dif u_w/\dif t$ become negative inside the star 
cluster volume. On the other hand, if the expansion velocity remains 
subsonic at $r > R_{sc}$, the right-hand side of equation
(\ref{eq.3a}) and the derivative of the expansion velocity, become 
negative outside the star forming region.

The above implies that a stationary wind solution, which assumes a 
continuous gas acceleration, exists only if the outflow crosses the 
star cluster surface at the local sound speed ($u = c_s$ at $r = R_{sc}$).
This conclusion, illustrated in Figure 1 doesn't depend on the 
wind thermodynamic properties. It is valid both for the adiabatic 
(CC85 and Cant{\'o} et al. 2000) and for the radiative solution.

There are three possible types of integral curve (Figure 1)
corresponding to different possible positions of the sonic point
with respect to the star cluster surface. 
1) The stationary wind solution (solid line, $R_{sonic} = R_{sc}$). 
In this case the thermal pressure decreases continually outside the star 
cluster surface, approaching a negligible value at large radii.
2) The breeze solution (dashed line, $R_{sonic} > R_{sc}$). In this
case the central temperature T$_c$ is smaller than in the first case,
which shifts the sonic point outside the cluster. This branch of solutions 
requires of a finite confining pressure and leads to zero velocity at 
infinity, in agreement with Parker's (1958) conclusion.
3) The unphysical double valued solution (dotted line, $R_{sonic} < R_{sc}$).
In this case T$_c$ is larger than in the stationary wind case. The thermal 
pressure goes to zero when the expansion velocity approaches the wind 
terminal speed.
%---------------------------------------------------------------
\begin{figure}[htbp]
\plotone{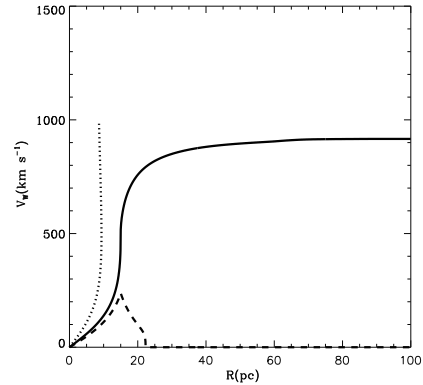}
\caption{Three possible types of integral curves. 1) The stationary 
wind solution (solid line), $R_{sonic} = R_{sc}$.  2) The breeze
solution (dashed line), $R_{sonic} > R_{sc}$. 3) Unphysical double 
valued solution (dotted line), $R_{sonic} < R_{sc}$.
For the three examples we adopted ${\dot E}_{sc}=10^{41}$ erg s$^{-1}$, 
$R_{sc}=15$ pc, $V_{\infty A} = (2 q_e/q_m)^{1/2} = 1000$ km s$^{-1}$.
\label{fig1}}
\end{figure}
%---------------------------------------------------------------

The appropriate solution is selected by the central conditions.
In order to obtain a stationary free wind solution, one has to find 
the wind central density and central temperature which 
accommodate the sonic point at the star cluster surface 
($u = c_s$ at $r = R_{sc}$). This is the key point that allows for the
definition of the central ($r$ = 0) wind parameters and for the
numerical solution of equations (\ref{eq.2a}-\ref{eq.3c}). 

The wind central temperature $T_c$ and central atomic number density 
$n_c$ are not independent in the radiative case. They are related 
by the equation
%---------------------------------------------------------------
\begin{equation}
      \label{eq.4}
n_c = \sqrt{\frac{q_e - \frac{q_m}{\gamma - 1}c^2_c}
       {\Lambda(T_c)}} = 
       q^{1/2}_m \sqrt{\frac{\frac{V^2_{\infty A}}{2} - 
       \frac{c^2_c}{\gamma - 1}}
       {\Lambda(T_c)}} .
\end{equation}
%---------------------------------------------------------------
This results from comparing the derivative 
of the expansion velocity at the star cluster center, using 
equation (\ref{eq.2c}), with equation (\ref{eq.2a}). 
Note that in the absence of radiative cooling, when
%---------------------------------------------------------------
\begin{equation}
       \label{eq.4b}
Q = n^2_c \Lambda(T_c) = q_e - \frac{q_m}{\gamma-1}c^2_c = 0 ,
\end{equation}
%---------------------------------------------------------------
equation (\ref{eq.4}) is transformed into the adiabatic relation
(\ref{eq.0c}). Thus, the wind parameters at the star cluster 
center can be found by iteration of the central temperature until 
the sonic point takes its proper position at the selected star 
cluster surface.

As in the adiabatic case $\rho_c$ and $T_c$ are independent,  
one can always find the central pressure that accommodates
the sonic point at the star cluster surface (see equations
\ref{eq.0a} and \ref{eq.0c}):
%---------------------------------------------------------------
\begin{equation}
      \label{eq.5}
R_{sonic} = R_{sc} = \frac{6 \gamma}{\gamma-1}\frac{B P_c}
                     {\sqrt{2 q_e q_m}} . 
\end{equation}
%--------------------------------------------------------------- 
However this is not the case if radiative cooling is taken into account.
In this case the central density and the central pressure go to zero
when the central temperature approaches its maximum
value $T_{max} = \frac{\gamma-1}{\gamma}\frac{\mu}{k}\frac{q_e}{q_m}$
(see equation \ref{eq.4}). $P_c$ increases for smaller
values of the central temperature. However, it cannot exceed a maximum 
value (see equation \ref{eq.4} and Figure 2a), bound by the gas 
radiative cooling. At this critical stage the fraction of energy 
radiated away at the cluster center  per unit time,
$\delta = (q_e-n^2_c\Lambda(T_c))/q_e$, reaches $\approx 30\%$
of the injected energy (see Figure 2c). If the central temperature 
becomes even smaller, there is no density enhancement able to 
compensate the fall in pressure promoted by radiative cooling. 
Consequently, the central pressure cannot promote an effective 
outward acceleration. Therefore in the radiative case, the sonic 
radius ($R_{sonic}$) cannot be arbitrarily large and has a maximum 
value for any given set of star cluster parameters. 

Figure 2b shows how $R_{sonic}$ depends on the central temperature, 
for particular values of $q_e$ and $q_m$. For the largest central 
temperature, the resultant central pressure acquires its lowest value
(see Figure 2a) and the sonic point lies very close to 
the star cluster center. Smaller values of $T_c$ lead to a larger $P_c$ 
(Figure 2a) and consequently to larger $R_{sonic}$ values. $R_{sonic}$ 
reaches its maximum possible value when the central pressure 
approaches also its maximum (compare figures 2a and 2b) as expected 
in the quasi-adiabatic regime (see equation \ref{eq.5}). 

%---------------------------------------------------------------
\begin{figure}[htbp]
\plotone{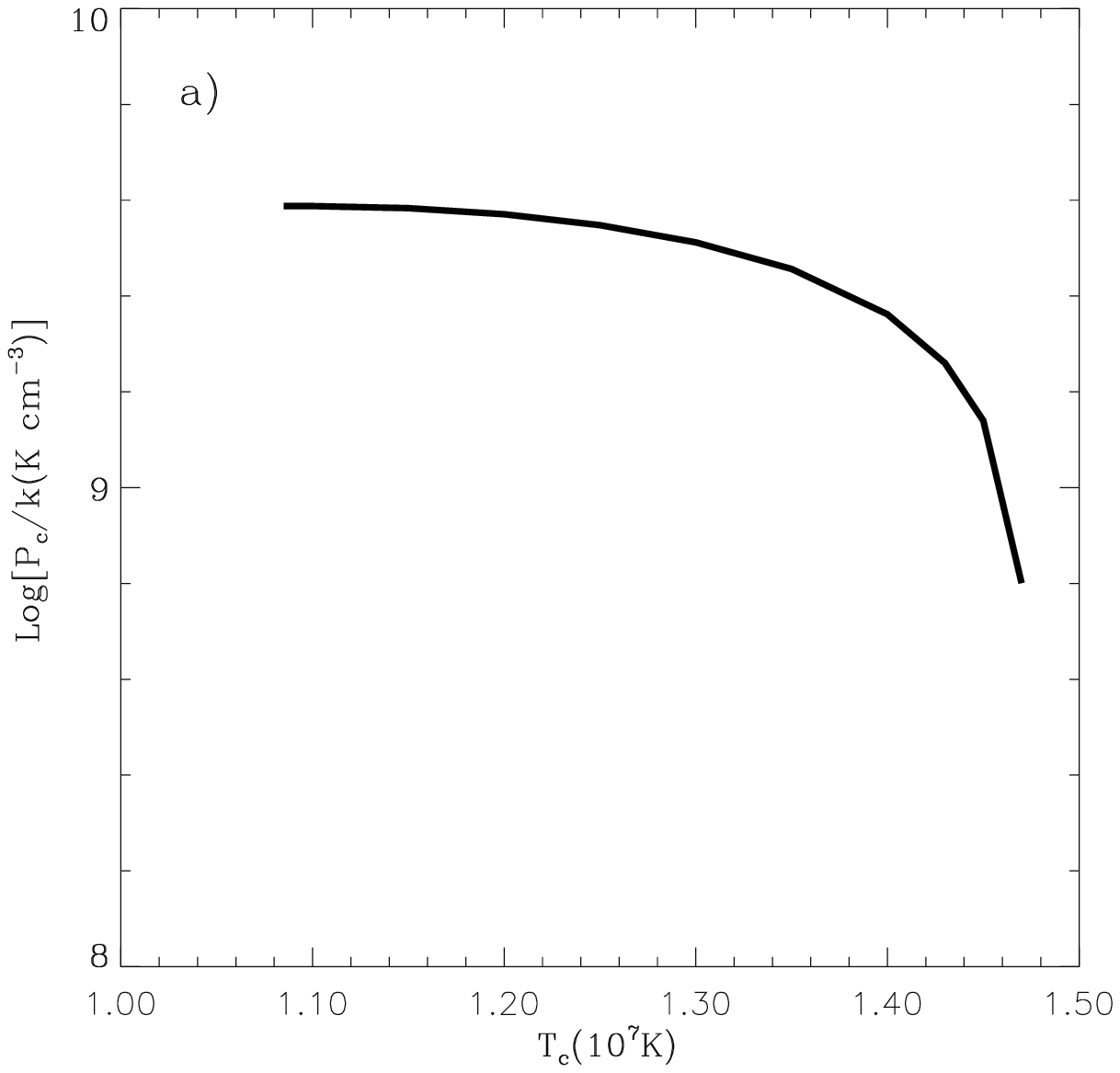}
\plotone{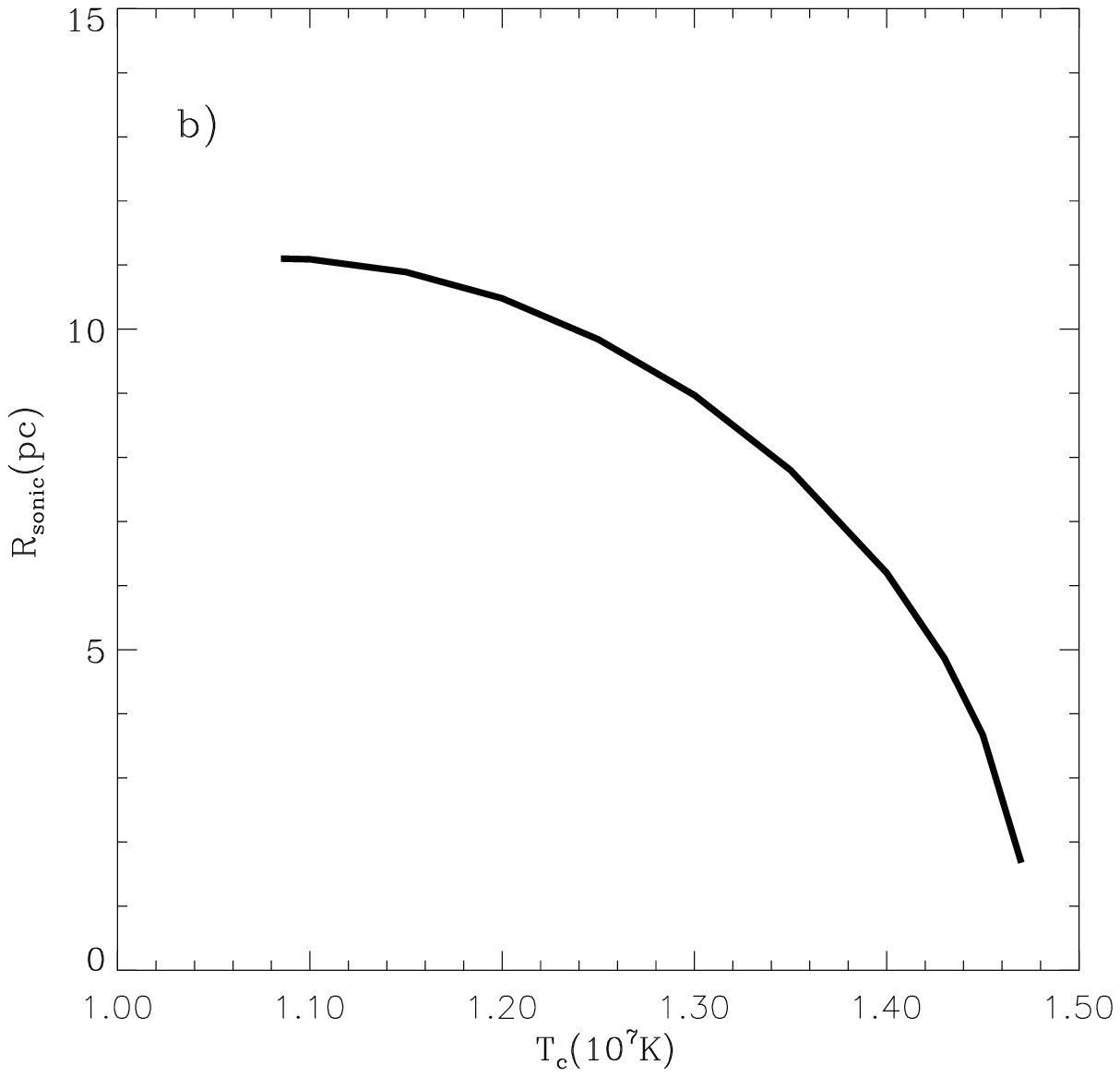}
\plotone{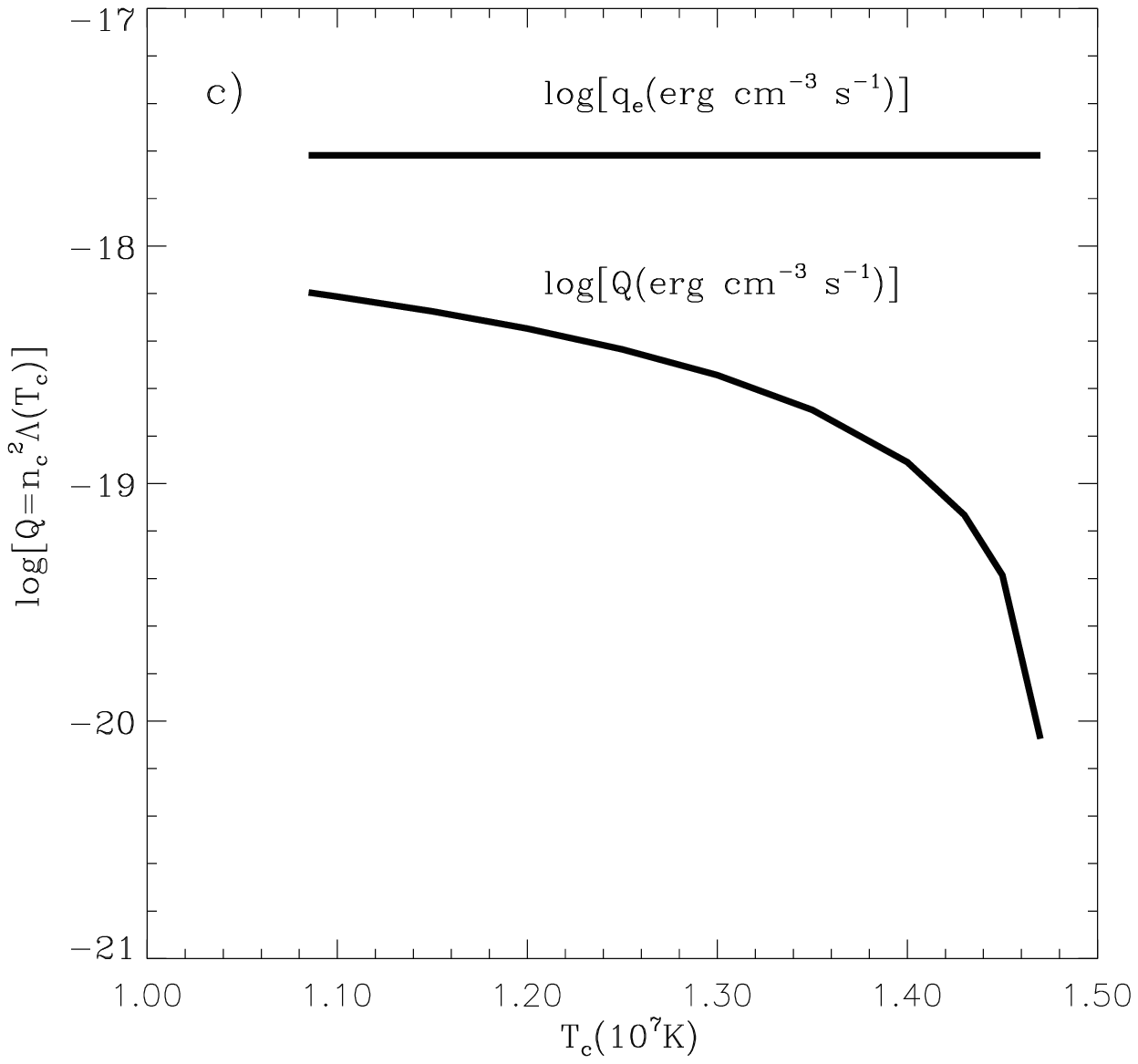}
\caption{The impact of the central temperature on the outflow.
a) The outflow central pressure; b) The position of the sonic 
point; c) Comparison of the deposited and the radiated energies
at the star cluster center. In all cases the energy and the mass 
deposition rates per unit volume are
$q_e = 2.4 \times 10^{-18}$ erg cm$^{-3}$ s$^{-1}$ and
$q_m = 4.8 \times 10^{-34}$ g cm$^{-3}$ s$^{-1}$, respectively.
$(2q_e/q_m)^{1/2}=V_{\infty,A}=1000$ km s$^{-1}$, and the assumed 
ejected gas metallicity is solar. 
\label{fig2}}
\end{figure}
%---------------------------------------------------------------

In the radiative stationary solution, although the sonic point may
approach its maximum possible value, radiative losses of energy would 
represent only a moderate fraction of the energy input rate. In these 
cases, cooling would drastically modify the wind temperature 
distribution outside the star forming region (see Paper I and 
discussion below).
When the rate of energy radiated away at the star cluster center exceeds 
$\sim 30$\% of the energy deposition rate, the stationary solution is
inhibited. This conclusion is stressed in Figure 3, which displays the 
critical energy deposition rate for different values of 
$q_e / q_m = V^2_{\infty A}/2$.  
%---------------------------------------------------------------
\begin{figure}[tb]
\plotone{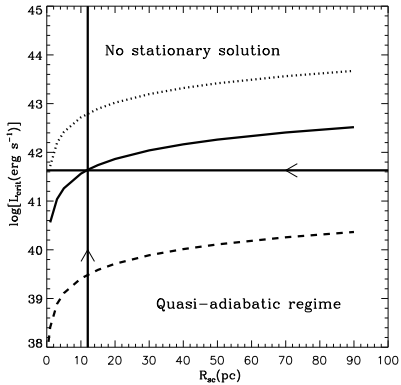}
\caption{The impact of radiative cooling. The threshold energy input 
rate above which the stationary wind solution is fully inhibited, as 
function of the star cluster radius. The solid line represents the 
threshold energy for star clusters with
$(2q_e/q_m)^{1/2}=V_{\infty A}=1000$ km s$^{-1}$. The dotted and 
the dashed lines mark the threshold energies 
for star clusters with $(2q_e/q_m)^{1/2} = 1500$ km s$^{-1}$ and 
500 km s$^{-1}$, respectively.
\label{fig3}}
\end{figure}
%---------------------------------------------------------------
Moving from right to left along the horizontal line is equivalent to
considering progressively more compact clusters, all with the same 
energy and mass deposition rates (${\dot E}_{sc} \approx 4.4 \times 
10^{41}$ erg s$^{-1}; {\dot M}_{sc} \approx 1.4$ \Msol \, yr$^{-1}$). 
For large star clusters the maximum allowed 
sonic radius $R_{sonic}$ exceeds the star cluster radius $R_{sc}$,
however one can accommodate the sonic point at the star cluster
surface once a proper central temperature is selected and obtain a 
stationary wind solution. However, if the considered star cluster is smaller 
than the critical value ($\sim 12$~pc for the example shown in Figure 3), 
the maximum allowed sonic point radius moves inside the star cluster 
and the stationary wind solution vanishes. 

The same is true if one moves along the vertical line in Figure 3, from low
to high energy input rates. In this case one is selecting progressively
more energetic star clusters within the same volume, until the sonic
point ends up inside the star cluster (in our example at $L_{crit} \approx 
4.4 \times 10^{41}$~erg s$^{-1}$) and the stationary wind solution vanishes. 
 
Once the proper initial conditions are selected, one can solve the
main equations (\ref{eq.2a}-\ref{eq.3c}) numerically and obtain
the wind temperature and density distributions. We have compared
for example our results with Stevens \& Hartwell (2003) standard model 
($R_{sc}=1$ pc, ${\dot M}_{sc}=10^{-4}$ \Msol yr$^{-1}$, $V_{\infty A}=
2000$ km s$^{-1}$). In this case the stationary wind evolves in 
the quasi-adiabatic regime and we found an excellent
agreement with Stevens \& Hartwell central values and X-ray 
luminosity. Our model predicts $T_c=5.9 \times 10^7$K, 
$n_c=0.65$ cm$^{-3}$ and the X-ray flux between 0.3 and 
8.0~keV from the central 1pc volume $L_x=5.2 \times 10^{32}$erg s$^{-1}$.

\section{The expected appearance of stationary radiative winds}

Free winds present a four zone structure (Silich et al. 2003): 
a star cluster region filled with a hot X-ray plasma, an adjacent 
X-ray halo with a decreasing temperature distribution, the line 
cooling zone and a region of recombined gas, exposed to the UV and 
soft X-ray radiation from the inner zones and to the UV photons
emitted by the star cluster itself.

%---------------------------------------------------------------
\begin{figure}[htbp]
\plotone{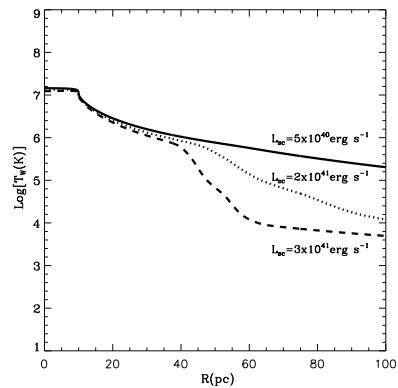}
\caption{The impact of cooling in the extended X-ray zone of
stationary winds.
Temperature profiles for progressively larger energy deposition
rates. Solid, dotted and dashed lines represent winds with 0.5, 2 
and 3 $\times 10^{41}$ erg s$^{-1}$, respectively. 
$R_{sc}=10$ pc in all three cases here considered.
\label{fig4}}
\end{figure}
%---------------------------------------------------------------
Figure 4 presents the free wind temperature distribution for three 
SSCs, all with the same $R_{sc}$ = 10 pc radius and  the same ratio
$(2q_e/q_m)^{1/2}=V_{\infty A}=1000$ km s$^{-1}$, but
different energy and mass deposition rates. 
The lowest energy case (solid line) lies well into the adiabatic regime.
In the other two cases however, the radiative cooling clearly 
modifies the internal wind structure bringing the boundary of the 
X-ray zone and the photoionized envelope closer to the star cluster 
surface. In the most energetic case shown in Figure 4, that with  
$3 \times 10^{41}$ erg s$^{-1}$ star cluster (dashed line), the outer 
boundary of
the X-ray zone ($T_{X-ray} \sim 5 \times 10^5$ K) is about a factor of 1.5 
smaller than in the adiabatic case. Furthermore, the dimension ($R_4$) 
at which the gas attains a temperature $\sim 10^4$ K lies about 10 
times closer to the star cluster center than in the adiabatic case. 
Consequently, the maximum density of the emission line 
envelope is $\sim 10^2$ times larger and the emission measure is 
$\sim 10^3$ times larger than in the adiabatic case. 

The line cooling zone and the photoionized envelope may be observed 
as a broad ($\sim 1000$ km s$^{-1}$) emission line component
perhaps of low intensity if compared to the narrow line  
caused by the central HII region. 

It is worth noticing that the luminosity of the central HII region
decays rapidly after $\sim 3$Myr, when the most massive stars in a
coeval cluster begin to move away from the main sequence, whereas the 
broad component conserves its luminosity being ionized by the soft X-ray 
radiation and therefore should be easier to detect in old 
($> 10^7$ yr) objects.

The X-ray luminosity of the star cluster wind is given by
%---------------------------------------------------------------
\begin{equation}
      \label{eq.6}
L_x = 4 \pi \int_0^{R_{x,cut}} r^2 n_w^2 \Lambda_x(Z_w,T_w) {\rm d}r ,
\end{equation}
%--------------------------------------------------------------- 
where $n_w(r)$ is the atomic density distribution, 
$R_{x,cut}$ is the X-ray cut-off radius where the wind temperature drops below 
$5 \times 10^5$K, and $\Lambda_x(Z,T)$ is the X-ray emissivity derived
by Raymond \& Smith (1977) in their hot-plasma code (see Strickland \&
Stevens 2000).

\section{Comparison with the observations}

\subsection{The Arches cluster} 

The Arches cluster is the densest and the most compact star cluster known 
in the Local Group. It is located within $\sim 0.2$pc volume at 
$\sim 50$ pc from the Milky Way center and contains $\sim 120$ 
stars with masses in excess of 20\Msol (Serabyn et al. 1998). The age of
the cluster is estimated within a range 2 - 4.5 Myr and the mass
is $\sim 10^4$\Msol \, for an IMF with $\alpha=1.6$ and 
lower and upper mass cutoffs of 1\Msol and 100\Msol, respectively.

Two sets of calculations for the Arches cluster wind have been
presented by Raga et al. (2001) and Stevens \& Hartwell (2003).
They differ somewhat on the assumed input parameters. Stevens \& 
Hartwell (2003) derived the total mass deposition rate ${\dot M}_{sc}=
7.3 \times 10^{-4}$\Msol \, yr$^{-1}$ and the average individual stellar 
wind terminal speed $V_{\infty} = 2810$ km s$^{-1}$ from the Lang et
al. (1999, 2001)
observations of the Arches cluster individual stars and adopted 
a Solar gas metallicity. This set of parameters leads to a total 
energy deposition rate ${\dot E}_{sc} \approx 1.8 \times 10^{39}$erg s$^{-1}$. 
Raga et al. (2001) assumed a lower mean individual stellar wind
terminal velocity ($V_{\infty} = 1500$ km s$^{-1}$) and presented
results for a star cluster with 60 identical massive stars, each ejecting
${\dot M}_{\star}= 10^{-4}$\Msol yr$^{-1}$. This implies a total
energy input rate ${\dot E}_{sc} \approx 4.2 \times 10^{39}$erg s$^{-1}$.
Note that the adiabatic wind X-ray luminosity between 0.3 and
8.0 keV for this set of parameters and 
$Z_w = 2 Z_{\odot}$ is $L_x \approx 3 \times 10^{37}$
erg s$^{-1}$, approximately two orders of magnitude above the
value indicated by Raga et al. (2001) 
($L_x \approx 3 \times 10^{35}$ erg s$^{-1}$). However we recover a good
agreement with their results if we adopt 
${\dot M}_{\star}= 10^{-5}$\Msol \, yr$^{-1}$ (instead of the cited value
${\dot M}_{\star}= 10^{-4}$\Msol \, yr$^{-1}$) for the individual
stellar mass loss rate. Fact that indicates a missprint throughout
their paper.

The results of the calculations that include radiative cooling for 
the modified (${\dot M}_{\star}= 10^{-5}$\Msol \, yr$^{-1}$)
Raga et al. (2001) star cluster model 
are presented in Figure 5a. For both sets of input parameters
(modified Raga et al., 2001 and Stevens \& Hartwell, 2003) the Arches
cluster wind evolves in both cases in the quasi-adiabatic regime. For the
Raga et al. (2001) star cluster parameters the central temperature 
is approximately $3.3 \times 10^7$K. It drops to the X-ray cut-off 
value ($5 \times 10^5$K) at 2.6 pc distance and to $10^4$K at 
46.2pc radius. The calculated 0.3-8.0keV X-ray luminosity is 
$L_x = 3 \times 10^{35}$erg s$^{-1}$, the broad emission line
luminosities are $L_{H\alpha} = 4.9 \times 10^{34}$erg s$^{-1}$,
$L_{Br\gamma} = 4.7 \times 10^{32}$erg s$^{-1}$, respectively.
For the Stevens \& Hartwell (2003) parameters 
(${\dot M}_{sc}= 7.3 \times 10^{-4}$\Msol yr$^{-1}$, 
$V_{\infty A} = 2810$ km s$^{-1}$) the central temperature
is $T_c \approx 1.2 \times 10^8$K, the  X-ray cut-off radius is 6.9 pc
and the inner boundary of the photoionized envelope (the $10^4$K radius)
is 129 pc. The calculated X-ray luminosity between 0.3 and 8.0 keV
and the broad emission line luminosities are $L_x = 10^{35}$erg s$^{-1}$,
$L_{H\alpha} = 6.6 \times 10^{34}$erg s$^{-1}$ and 
$L_{Br\gamma} = 6.3 \times 10^{32}$erg s$^{-1}$, respectively.
%---------------------------------------------------------------
\begin{figure}[htbp]
\plotone{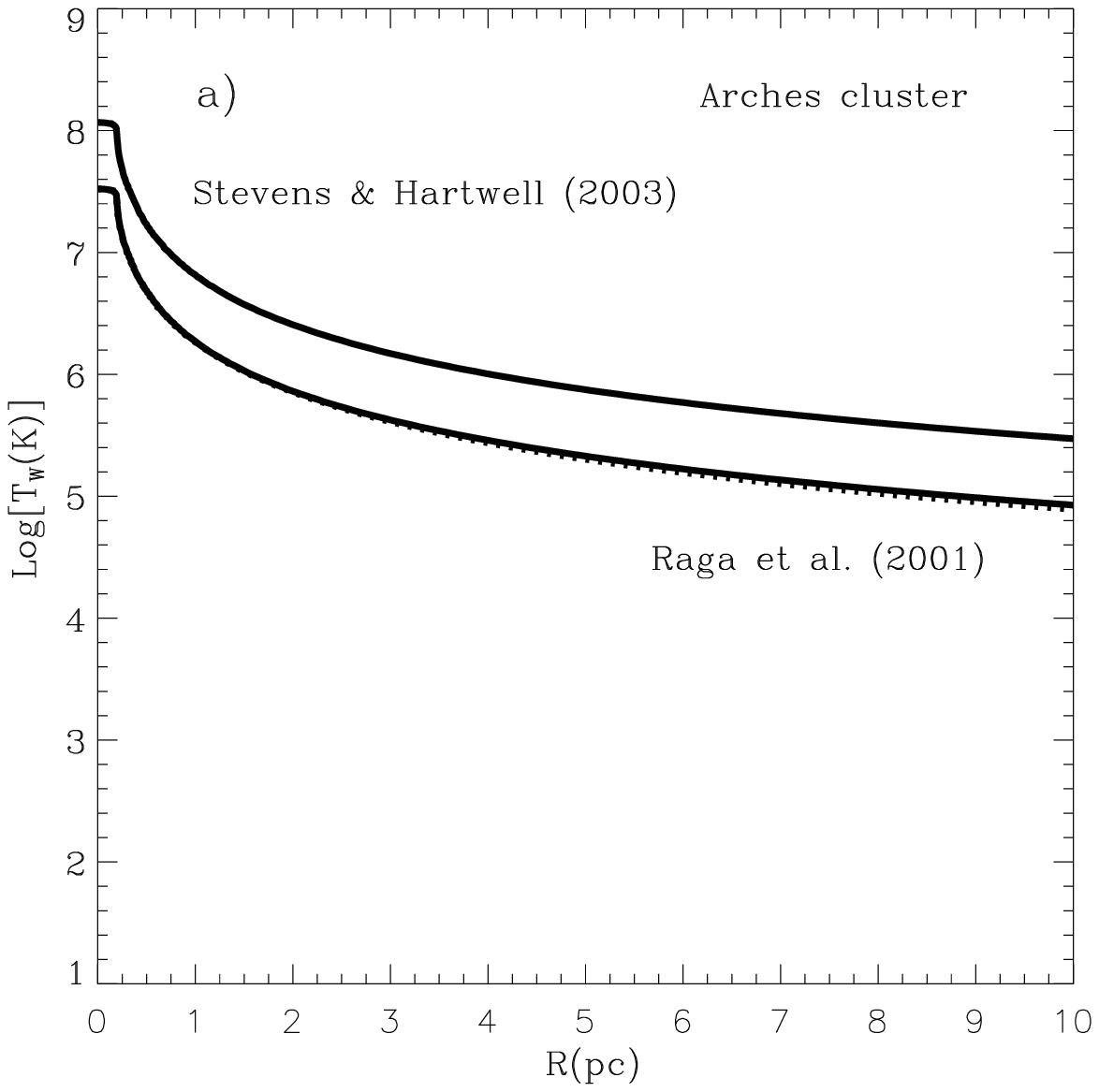}
\plotone{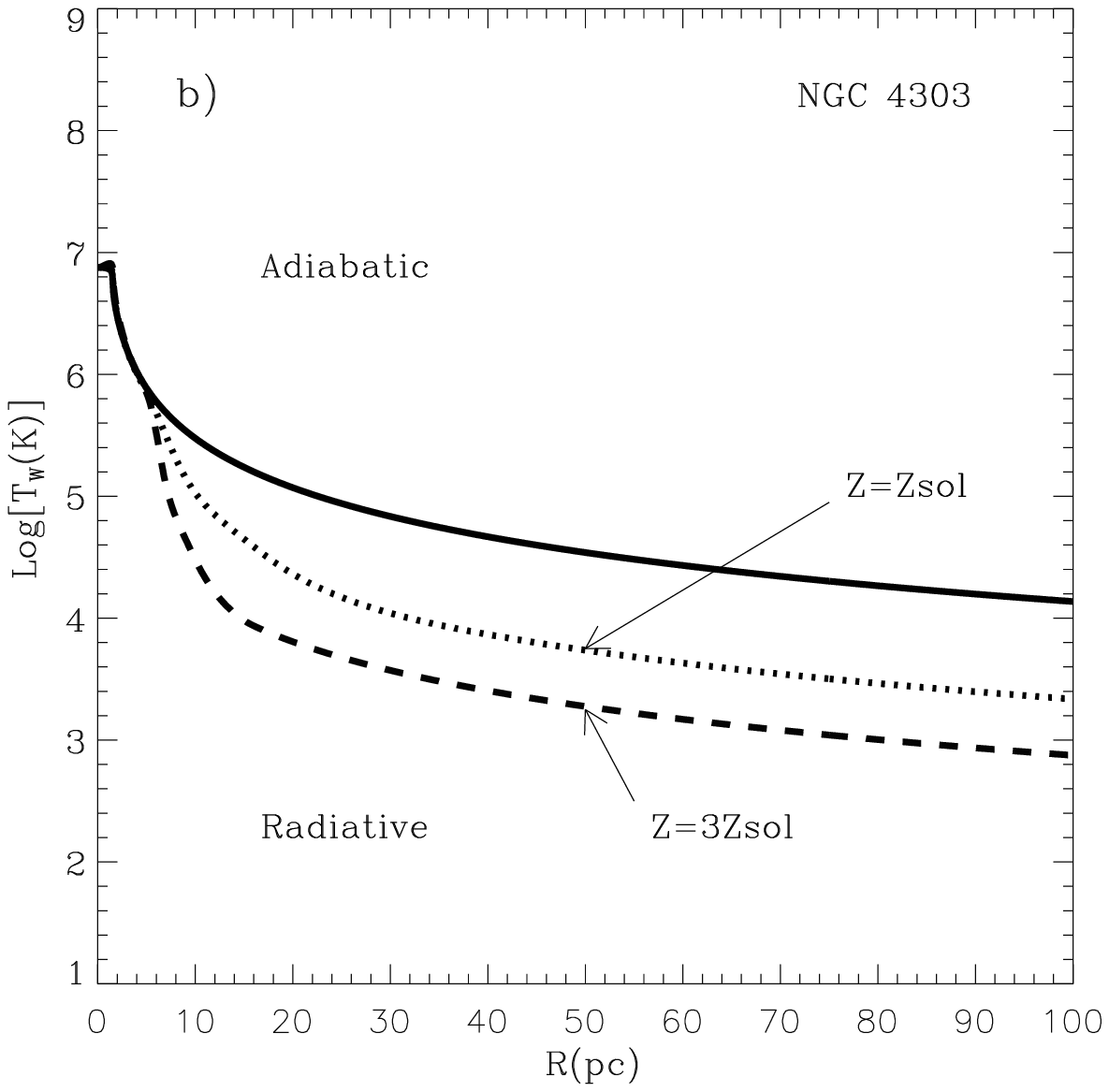}
\plotone{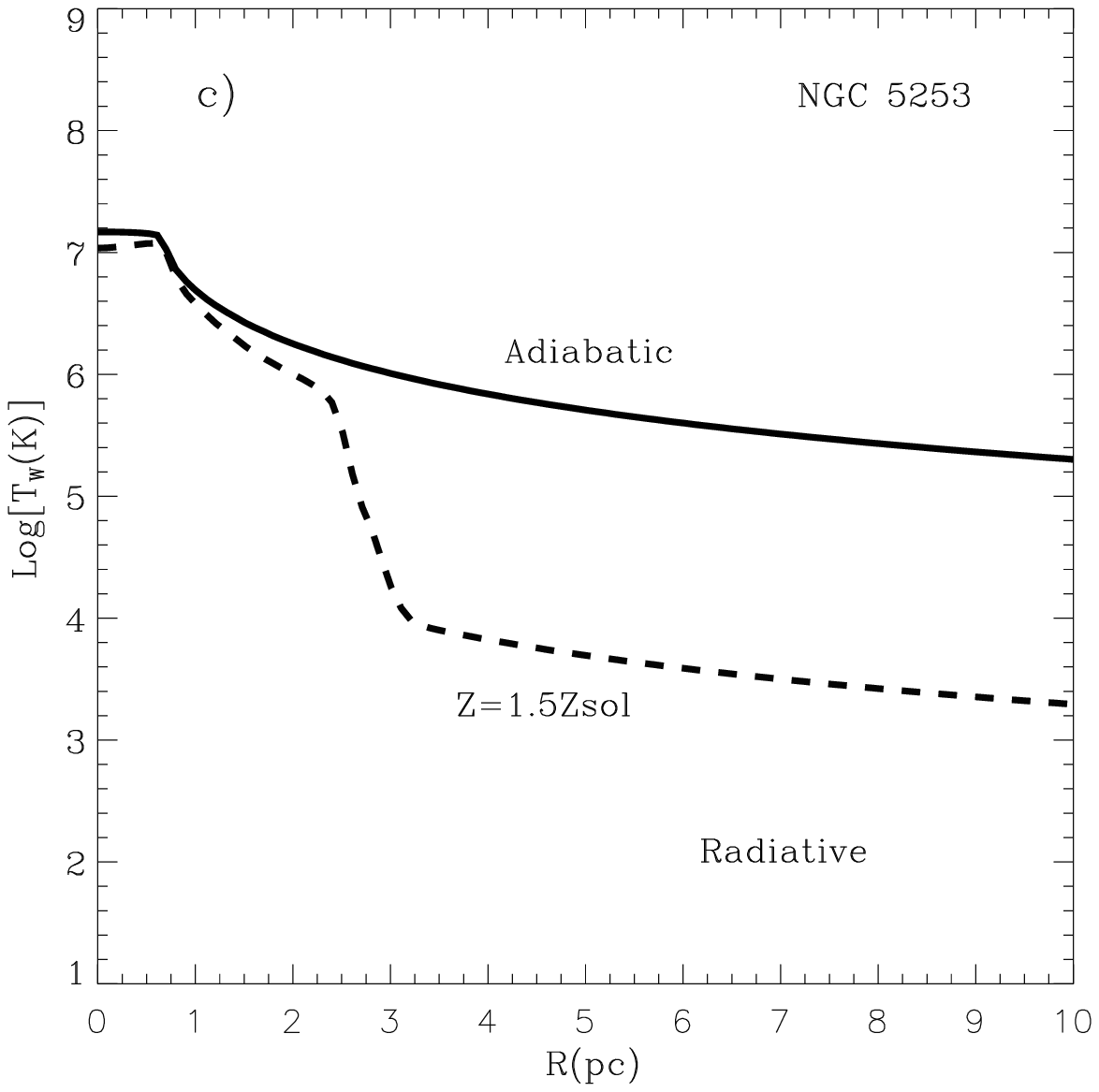}
\caption{The stationary wind temperature distributions.
a) The Arches cluster. Solid lines are Stevens \& Hartwell (2003)
and modified Raga et al. (2001) solutions. The dotted lines present
radiative solutions for the same sets of the star cluster parameters.  
b) The NGC 4303 nuclear super-star cluster. 
The adiabatic temperature distribution is shown by the solid line.
The dotted and dashed lines display the radiative model predictions
for solar and 3 times solar wind metallicities, respectively. 
c) The temperature distribution for NGC 5253 supernebula critical 
outflow. The adiabatic stationary wind solution (solid line) is 
compared with the radiative solution when $Z_w = 1.5Z_{\odot}$.
For larger values of $Z_w$ and the adopted cluster parameters the
stationary solution does not exist.
\label{fig5}}
\end{figure}
%--------------------------------------------------------------- 

\subsection{The nucleus of NGC 4303}

The galaxy nucleus of the NGC 4303 belongs to the class of low-luminosity
active galactic nuclei. The energy output rate from the nuclear
region is dominated by the compact ($R_{sc} \approx 1.55$ pc) and 
massive (M$_{sc} \approx 10^5$ \Msol) super-star cluster.  
The H$\alpha$ luminosity derived from the WHT/ISIS H$\alpha$ flux 
corrected for the absorption and aperture effects is
$1.2 \times 10^{39}$ erg s$^{-1}$ (Colina et al. 2002).
The SSC ultraviolet spectrum is best fitted by a 4 Myr old
instantaneous starburst of $10^5$\Msol \, with a Salpeter 
IMF and 1\Msol \, and 100\Msol \, lower and upper stellar mass cutoffs.
The thermal component of the unresolved-core X-ray spectrum 
is best fitted by T$ \approx 7.5 \times 10^6$K plasma with the 
X-ray luminosity between 0.07 keV and 2.4 keV around 
$2 \times 10^{38}$ erg s$^{-1}$ (Jim{\'e}nez-Bail{\'o}n et al. 2003).

We adopt a SSC mechanical luminosity, L$_{sc} \approx 3
\times 10^{39}$ erg s$^{-1}$, predicted by the Leitherer et al. (1999)
starburst model and associate the observed hot plasma temperature with
the wind central temperature and derive the wind terminal velocity
by iterations, from the condition that the flow crosses the star cluster
surface with the local sound speed. The calculated wind terminal velocity
is 715 km s$^{-1}$. Figure 5b present the temperature distribution
for this strongly radiative wind and shows how it begins to deviate 
from the adiabatic profile (solid line) at a distance $\sim 6$ pc 
away from the center. It falls to the X-ray cutoff value at 5.9 pc and 
reaches 10$^4$K value at 31.9 pc. The calculated X-ray luminosity for 
0.3 and 2.0 keV energy range and broad emission line 
luminosities are $L_x = 1.3 \times 10^{38}$erg s$^{-1}$,
$L_{H\alpha} = 1.5 \times 10^{36}$erg s$^{-1}$ and 
$L_{Br\gamma} = 1.4 \times 10^{34}$erg s$^{-1}$, respectively.
This implies that the expected H$\alpha$ broad 
luminosity constitutes only 0.1\% of the NGC 4303 core H$\alpha$
emission.

Figure 5b also demonstrates how the temperature profiles are sensitive
to the adopted ejected gas metallicity. In the case of 
$Z_w = 3 Z_{\odot}$ the X-ray cut-off radis is 5.6 pc, the temperature 
drops to the 10$^4$K value already at 14.6 pc and the calculated 
wind luminosities are $L_x = 2.6 \times 10^{38}$erg s$^{-1}$,
$L_{H\alpha} = 1.3 \times 10^{36}$erg s$^{-1}$ and 
$L_{Br\gamma} = 1.3 \times 10^{34}$erg s$^{-1}$, respectively.

\subsection{The NGC 5253 supernebula}

NGC 5253 is a nearby (3.8 Mpc) peculiar dwarf galaxy containing
numerous massive super-star clusters (Meurer et al. 1995; Gorjian
1996). An extraordinary compact and  bright radio-infrared
super-nebula has been discovered within the central star forming 
region of this galaxy by Turner et al. (2000) and Gorjian et al. 
(2001). The Lyman continuum rate required to maintain the ionization 
of the super-nebula, $N_{Lyc}=4 \times 10^{52}$ s$^{-1}$, requires 
of a $(5-7) \times 10^5$\Msol star cluster with energy deposition rate 
${\dot E}_{sc} \approx 2 \times 10^{40}$ erg s$^{-1}$ and
a Salpeter mass distribution having 100\Msol and 1\Msol mass cut-off 
limits (see Leitherer et al. 1999). The mean radius of the ionizing 
cluster is $R_{sc} \approx 0.7$ pc. Using the Keck Telescope recombination
line spectra, Turner et al. (2003) obtained a recombination 
linewidth of 75 km s$^{-1}$ and concluded that the super-nebula gas
may actually be bound by the gravitational pull of the super-star
cluster. 

It is worth noticing that the above parameters imply that NGC 5253
super-star cluster is close to our critical energy limit 
(see Figure 3) if one assumes that the energy to mass deposition rates 
ratio is close to the $(2q_e/q_m)^{1/2} \sim V_{\infty,A}=1000$ km 
s$^{-1}$ (solid line in Figure 3). If this is the case a small
increase in the ejected gas metallicity (to more than $Z_w = 1.5Z_{\odot}$) 
will move star cluster above the critical value into the forbidden
parameter space. This implies that the stationary wind solution may 
not exist for this particular cluster and that the ejected gas is 
to accumulate in the neighborhood of the star cluster. In such a case, 
the super-nebula may consist of thermalized matter injected by winds 
and supernovae, which under strong radiative cooling acquires a low 
sound speed value and is thus unable to stream at high velocities 
away from the cluster as a stationary wind.

\section{Conclusions}

We have developed a self-consistent stationary solution for 
spherically symmetric winds driven by compact star clusters 
taking into consideration radiative cooling. 

We have shown that stationary radiative winds differ strongly from
their adiabatic counterparts. In particular we have shown 
that in the energy-size plane, there is a regime where
the stationary wind solution is inhibited. This occurs whenever
the energy radiated away per unit volume and per unit time
($n^2_c \Lambda(Z_w,T_w)$) surpasses a value of $\sim 30\%$ of the energy
injection rate. In this catastrophic cooling regime, the sonic point 
cannot be accommodated at the star cluster surface and the stationary 
wind solution does not exist. Below such a limit the flow, despite 
radiative cooling, behaves within the star cluster volume in a 
quasi-adiabatic manner and is able to set the sonic point at the 
star cluster boundary and evolve into a stationary wind. 

Stationary winds driven by stellar clusters with an energy input rate or a
size that approaches the critical value, establish a temperature 
distribution radically different from that predicted by the adiabatic 
solution. In these stationary wind cases the fast fall of temperature 
brings the boundaries of the X-ray zone, and of the line cooling zone 
and the photoionized envelope, closer to the star cluster center. This
promotes the establishment of a compact ionized gaseous envelope 
which should be detected as a week and broad ($\sim 1000$ km s$^{-1}$) 
emission line component at the base of a much narrower line caused by 
the central HII region.

Note that the threshold energy input rate approaches an asymptotic
value for large values of $R_{sc}$ (see Figure 3). This implies
that single supermassive star clusters are not able to generate
stationary outflows whatever their radii may be. The fate of the 
ejected material in this case remains unclear. 
A self-regulating star forming region may form and may keep 
the injected gas bound because of catastrophic radiative cooling
or the gas may be blown away in a quasi-recurrent regime.
The outflows driven by supermassive or super-compact star clusters
should be studied with a full non-stationary hydrodynamic approach.

We have speculated that the super-nebula in NGC 5253 seems a good
example of this inhibited stationary wind regime. Radiative cooling
enforces a rapid drop in the sound speed value and the injected matter
is to remain near the cluster. In such a case we predict that the
metallicity of the super-nebula is above solar, making the cluster lie
above the threshold limit for stationary winds (Figures 3 and 5).

Our calculations show that the Arches cluster wind seems to
evolve in the quasi-adiabatic regime and predict the H$_{\alpha}$ 
and Br$\gamma$ broad component luminosities around
$L_{H\alpha} \approx 5 \times 10^{34}$ erg s$^{-1}$ and
$L_{Br\gamma} \approx 5 \times 10^{32}$ erg s$^{-1}$, respectively.

The temperature distribution derived for the NGC 4303 central
1.55~pc star cluster wind is radically different from the 
adiabatic temperature distribution even for a solar wind metallicity.
The calculated X-ray luminosity is in reasonable agreement with
the observed diffuse component luminosity. The radiative model
also predicts a compact (between 6 pc and 30 pc) broad line
emission with $L_{H\alpha} \approx 10^{36}$ erg s$^{-1}$
and L$_{Br\gamma} \approx 10^{34}$ erg s$^{-1}$, 
respectively. 

We are pleased to thank D. Strickland who provided us with his
X-ray emissivity tables. Our thanks also to C. Mu\~noz-Tu\~n\'on 
for multiple suggestions and to C. Law for a useful discussion about 
the Arches cluster observational parameters during the X-ray - radio 
connection Santa Fe workshop. We thank prof. J. Palou\v s for his
comments and suggestions as a referee. We also appreciate the
financial support given by M\'exico (CONACYT) research grant 36132-E.

\newpage

%\onecolumn


%\begin{figure}
%\figcaption[ms15457_1.ps]
%{The dispersion relation $\omega (\eta )$ for $c_{sh}$ corresponding to 
%position of a 100 M$_{\odot}$ fragment. \label{fig1}}
%\end{figure}

\end{document}